\documentclass[a4paper]{IEEEtran}
\usepackage{graphicx}
\usepackage{url}
\usepackage{xspace}
\usepackage{subfig}

\newcommand{\tool}{{\tt tw}{\sf Awler}\xspace}

\newcommand{\feat}{\sf}

\begin{document}

\title{\tool: A lightweight twitter crawler}
\author{
  \IEEEauthorblockN{Polyvios Pratikakis} \\
  \IEEEauthorblockA{
    FORTH-ICS\\
    polyvios@ics.forth.gr
  }
}
%\affiliation{%
%  \institution{Institute of Computer Science, FORTH}
%  \city{Heraklion}
%  \state{Crete}
%}
%\email{polyvios@ics.forth.gr}

\maketitle

\begin{abstract}

This paper presents \tool, a lightweight twitter crawler that targets
language-specific communities of users. \tool takes advantage of
multiple endpoints of the twitter API to explore user relations and
quickly recognize users belonging to the targetted set.  It performs a
complete crawl for all users, discovering many standard user
relations, including the retweet graph, mention graph, reply graph,
quote graph, follow graph, etc.  \tool respects all twitter policies
and rate limits, while able to monitor large communities of active
users.

\tool was used between August 2016 and March 2018 to generate an
extensive dataset of close to all Greek-speaking twitter accounts
(about 330 thousand) and their tweets and relations.  In total, the
crawler has gathered 750 million tweets of which 424 million are in
Greek; 750 million follow relations; information about 300 thousand
lists, their members (119 million member relations) and subscribers
(27 thousand subscription relations); 705 thousand trending topics;
information on 52 million users in total of which 292 thousand have
been since suspended, 141 thousand have deleted their account, and 3.5
million are protected and cannot be crawled.  
\tool mines the collected tweets for the retweet, quote, reply, and
mention graphs, which, in addition to the follow relation crawled,
offer vast opportunities for analysis and further research.

\end{abstract}

\section{Introduction}

Social media content offers many opportunities for research in
numerous fields and disciplines, including machine learning, natural
language processing, epidemiology, sociology, economics, etc.  Data
mining social media is, however, increasingly difficult, due to
technical and policy constraints.

Specifically, twitter has been
used in multiple studies and analyses due to its more open and public
content.  Twitter restricts the reuse and publication of data crawled
using its public API to only sharing anonymized information (user and
tweet IDs), and moreover restricts the number of queries made to its
API to a limited rate.  This limitation may be overcome by combining
multiple users' rate limits and aggregating multiple crawls, but (i)
this may be considered sharing non-anonymized information by twitter,
or (ii) it may require access to expensive resources such as several
cloud VMs or a cluster, for a prolonged period of time.

This paper presents \tool, an open-source\footnote{The software will
be released under Apache License before this publication is in
print. \tool is about 11KLoC of Python.}, cost-effective, lightweight twitter crawler that can explore, discover and target
users related to a specific topic or using a given language.  The
crawler takes advantage of multiple twitter API endpoints, maximizing
the total crawled information while respecting all limits and
policies.  Moreover, it requires a single users' credentials
and can run on a single machine over large periods of time, tolerating
reboots and downtime without issue.  \tool aims to be complete,
\emph{i.e.}, does not crawl a sample of the user content but instead all of
the traffic of the users belonging to the crawled community.  The
crawler can analyze tweets and generate a multitude of relations and
information, including the follow, retweet, mention, reply, quote, and
favorite graphs, temporal patterns, topic and word frequencies, etc.  The
author has used the crawler for a period of 20 months to
discover and track a set of all Greek-speaking twitter accounts, using
a low-cost machine, the author's desktop PC.  \tool can perform
similarly well for even larger twitter communities, especially when
targetting language-specific parts of the twitter graph.

%The remainder of this paper is organized as follows; Section

\section{Twitter Crawler}

\tool is a custom crawler for twitter data that discovers and monitors
Greek-speaking twitter users, monitors all their publicly accessible
content.  The crawler stores this information and is able to extract
multiple relations, including the follow graph, the mention, reply,
retweet and quote graphs, the favorited graph, etc.  These relations
are timestamped, enabling further analysis using dynamic graph
techniques.  \tool maintains a set of tracked users, a set of users
that have been marked as greek-speaking, a set of stop-users that are
definitely not greek-speaking and the sets of dead, suspended, and
protected users. \tool is structured as a set of small tools and
the scripts using them to perform on-demand or continuous crawls.

\subsection{Tweet Crawler}
\tool uses the twitter \textsf{/statuses/user\_timeline} API endpoint
to crawl the tweets of tracked users.
To crawl the selected users' tweets, \tool downloads all the user
tweets that it can using the request twitter API, up to the 3200
tweets that twitter allows, or until it reaches the last tweet seen
when the user was crawled previously.
To save on the number of rate limited requests, \tool
prioritizes crawling of users that have a high probability of having
tweeted a lot since last crawled.  To do that, it computes the average
tweets per day for every user and uses two processes of tweet
crawling:
On the one hand, the crawler sorts all users based on their expected
tweets since they were last crawled and crawls users with a high
number of expected tweets.
On the other hand, the crawler sorts all users based on the time since
they were last crawled and crawls users not visited in a long time.
This way, \tool minimizes the probability of lost tweets, and also
make the best use of the available requests per minute that twitter
allows.
Note that twitter throttles the number of API requests per 15 minutes,
that only the last 3200 tweets can be requested, and that a maximum of
200 tweets can be returned per request.
A simple back-of-the-envelope calculation shows that by delaying
crawling to keep the average number of tweets per user crawled close
to 1000 since the last visit, we consume half the number of requests
per tweet compared to an average of 100 tweets since the last visit.

In addition to the stardard twitter user timeline crawling, \tool also
looks for tweets from tracked users that are retweets, reply to, or
quote other tweets, and use the \textsf{/statuses/lookup} endpoint to
ask specifically for information regarding these tweets and their
authors.  This way, the crawler will eventually discover all
interactions of the tracked users and crawl them for analysis of the
whole ``thread''.

To discover new users in the targetted community, we seed the tracked
set of users by performing a search for common greek stopwords on the
\textsf{/statuses/filter} twitter streaming API.  There is no need to
run the filter continuously, as it is only used to seed the tracked
users, since the crawler is user-centric and aims for a full crawl of
the involved users.
In addition to seeding based on stopwords, the crawler adds new users
by tracking any retweets in Greek from currently tracked users.
Specifically, if the crawler discovers 10 tweets in Greek by an
unknown user that have been retweeted by tracked users, then it also
starts tracking the new user as there is a high probability they are
also Greek-speaking.
Moreover, a daily pass discovers tracked users with more than 500
tweets, of which less than 2 per cent are in Greek, and stops tracking
them, adding them to the set of users where the crawl stops, as they
were found to not be greek-speakers.

\subsection{Follow Graph Crawler}

The public twitter API offers four ways of crawling user follow
relations, namely using the \textsf{friends/ids} and
\textsf{friends/list} API endpoints for crawling the IDs or fully
populated user structs of the user's friends, and
\textsf{followers/ids} and \textsf{followers/list} to crawl followers
respectively.  As rate limits are set separately for each endpoint,
\tool crawls all users' friends and followers using four crawler
processes, marking each crawled user to not be crawled again for the
following 30 days.  This limit is currently arbitrary, as the follow
relation does not change often or oscillate, and could be set to a
much lower time window while still remaining within the rate limits
for a community of this size.  Note that two of the four endpoints
\tool uses to crawl the follow graph produce detailed user
information, whereas the other two produce solely user IDs.  These are
easily separated and stored as the follow graph and a table of users,
where some IDs may not yet be populated with all user information.
\tool periodically runs a separate processes to populate missing user
information, using the \textsf{/users/show/:id} endpoint, and also
utilize the tweet crawler to save the user information for any user
for whom it finds it missing or out-of-date.  Note that this will
generate multiple entries of the user information per user, allowing
an analysis to monitor the evolution of the user's profile, bio,
language, etc, over time.  Currently, \tool is configured to use a
time window of two weeks for crawling user information, and do not
store the new information if it only differs in the number of tweets
from the user profile last seen.

\subsection{List Crawler}

Twitter lists have been used to mine user-curated information in
previous research.  As they aggregate the opinion, or classification,
of users into groups by other users, they amount to valuable
crowdsourced information that can be mined to infer common interests,
relations, etc.  The twitter API offers 4 ways to crawl lists;
namely, \textsf{lists/subscriptions} returns information regarding
the lists to which a user subscribes, \textsf{lists/memberships} 
returns information regarding the lists of which a user is a member,
\textsf{lists/members} returns the users that are members of a
list, and \textsf{lists/ownerships} returns lists curated by the
given user.  As these have different rate limits, \tool uses four
separate crawler processes that crawl lists and users in a round-robin
fashion and populate or update the list membership relations.

\subsection{Favorites Crawler}

Twitter allows users to signal interest in a tweet by marking it as
``favorite'' (or ``like'').  The twitter API offers one way to get a
user's favorites, \textsf{favorites/list}, returning ``likes'' ordered
by date of ``liked'' tweet. This is one of the most limiting constraints,
as any ``likes'' on tweets older than the newest ``like'' seen will be
lost.  To avoid that as much as possible, \tool does not stop after
observing previously seen ``likes'' as with tweets, but continues
crawling until having seen more than 190 previously known ``likes''.
The probability of missing user ``likes'' remains, but as users often
``like'' tweets that they observe in the top of their timeline,
missing old ``likes'' happens less often.  The twitter web client
allows viewing a user's ``likes'' in the order they were done, not the
order they were tweeted; manually observing the ``likes'' of five
very active users over a period of two days showed a small percentage
of missed ``likes''.  \tool could scrape the web page for these, but
does not, so as to remain very clearly within the limitations of the
twitter API agreement.  Note that the rate limiting for favorite
crawling is an order of magnitude less than the tweet API limits,
resulting in only a sample of the favoriting graph compared to the
near-full coverage for the other information.

\subsection{Crawler Storage}

\tool currently uses a MongoDB installation to store crawled data.
The MongoDB contains the following collections, used to store the
crawled data as well as metadata used by the crawler.

The \textsf{users} collection contains user objects as
returned by the Twitter API. There may be multiple entries per single
user, as discussed above. Each entry is annotated with the insertion
date. To avoid sorting and aggregation when scanning for the latest
entry for a given user, the data includes an additional field called
\textsf{screen\_name\_lower} that contains the screen name in
lowercase.  This field is unique when it exists, and in aggregate
helps depict the currently existing users as last seen by the crawler.

The \textsf{tweets} collection contains tweet objects as returned by
the Twitter API.  Each tweet is unique.  As there may be missing or
truncated tweets in the data store, \tool uses additional curation
processes that filter out truncated tweets and use the
\textsf{/statuses/lookup} API to properly populate them with the
non-truncated version.  The author has used this tool to also properly
ingest textual dumps of tweets generated by early versions of the
crawler, and handle the switch of the tweet length limit from 140 to
280 characters seamlessly.

The \textsf{trends} collection contains trend objects as
returned by the \textsf{trends/place} Twitter API for Greece,
timestamped and crawled every 15 minutes.  The rate
limit for trends is never reached, as Greek-speaking accounts tend to
have a very large geographical correlation with Greece, and there is
currently no need to crawl trending topics for multiple locations.

The \textsf{shorturl} collection contains key-value pairs of shortened
URLs and their corresponding target URL, as resolved by \tool.  To
populate this collection, \tool uses both the expanded URL provided
for each shortened URL by twitter within each tweet, and also uses a
set of crawler processes that interact with URL shortener services or
directly with the web, to resolve shortened URLs.%  To drive these
%crawlers, \tool uses a list of URL shorteners assembled and curated by
%the author.

The \textsf{follow} collection stores directed, timestamped follow
edges as generated by the follow graph crawler.  This amounts to a
dynamic graph that approximates the follow relations between tracked
users and their friends and followers, \emph{i.e.}, it is a superset
of the follow relation between tracked users, and includes their
friends and followers regardless of whether their tweets are being
tracked.

The \textsf{greeks} collection contains IDs (key) and screen names
(not necessarily up-to-date) of users that have been classified as
Greek-speaking.  \tool is parametric as to the criterion; the
reported deployment is configured to classify users as Greek speakers
and track when they satisfy any of the following conditions:
\begin{itemize}
  \item Users with more than 100 tweets, of which at least 20\% are in Greek.
  \item Users with more than 500 tweets, of which at least 10\% are
  in Greek and their username and bio are in a set of common Greek
  names or written in the Greek alphabet.
\end{itemize}
Clearly, the Greek language having a unique alphabet aids
significantly in recognizing Greek speakers. However, that is not
central to \tool, as Twitter's language recognition is very precise in
several languages that use the latin alphabet.
Conversely, \tool marks users as non-Greek speakers and stops the
crawler from following them and discovering new users through them,
when it has crawled more than 500 of their tweets, of which less than
1\% are in Greek.
The author has found that these conditions succeed in classifying most
users into either Greek-speaking or not, leaving only a small number
of inconclusive users.  \tool applies an additional
constraint to these, where if more than 30\% of a user's
followers and friends have already been classified as Greek-speaking,
the inconclusive user is marked as Greek-speaking.

The \textsf{suspended} collection contains IDs and screen names of
users that have been observed to be suspended by twitter.  Screen
names may not necessarily be up-to-date, as the account may have
changed its screen name between the time it was last crawled and when
it was suspended.

The \textsf{ignored} collection contains the IDs of users that have
been dropped by the crawler using the criteria described above, or by
manual curation.

The \textsf{cemetery} collection contains IDs (key) and screen names
of users that have been seen to have deleted their account.  Note that
these users may reactivate their accounts at some point, at which they
may be re-discovered and removed from the deleted user's collection.
This is done only using user IDs and not screen names, as the latter
may change or be taken by a different user in the mean time.

The \textsf{crawlerdata} collection contains enough information for
the crawler to properly track the user without wasting any API
requests.  Namely, it contains the ID of the first and last tweets
seen by the user,  the earliest and latest times they have been
crawled, whether their API limit of 3200 tweets in the past has been
reached and thus only new tweets can be crawled, when the user profile
and avatar picture was last downloaded, and when the user's favorites
were last scanned.

\section{Post-processing and Vectorization}

\begin{figure}
\centering
\includegraphics[width=0.49\textwidth]{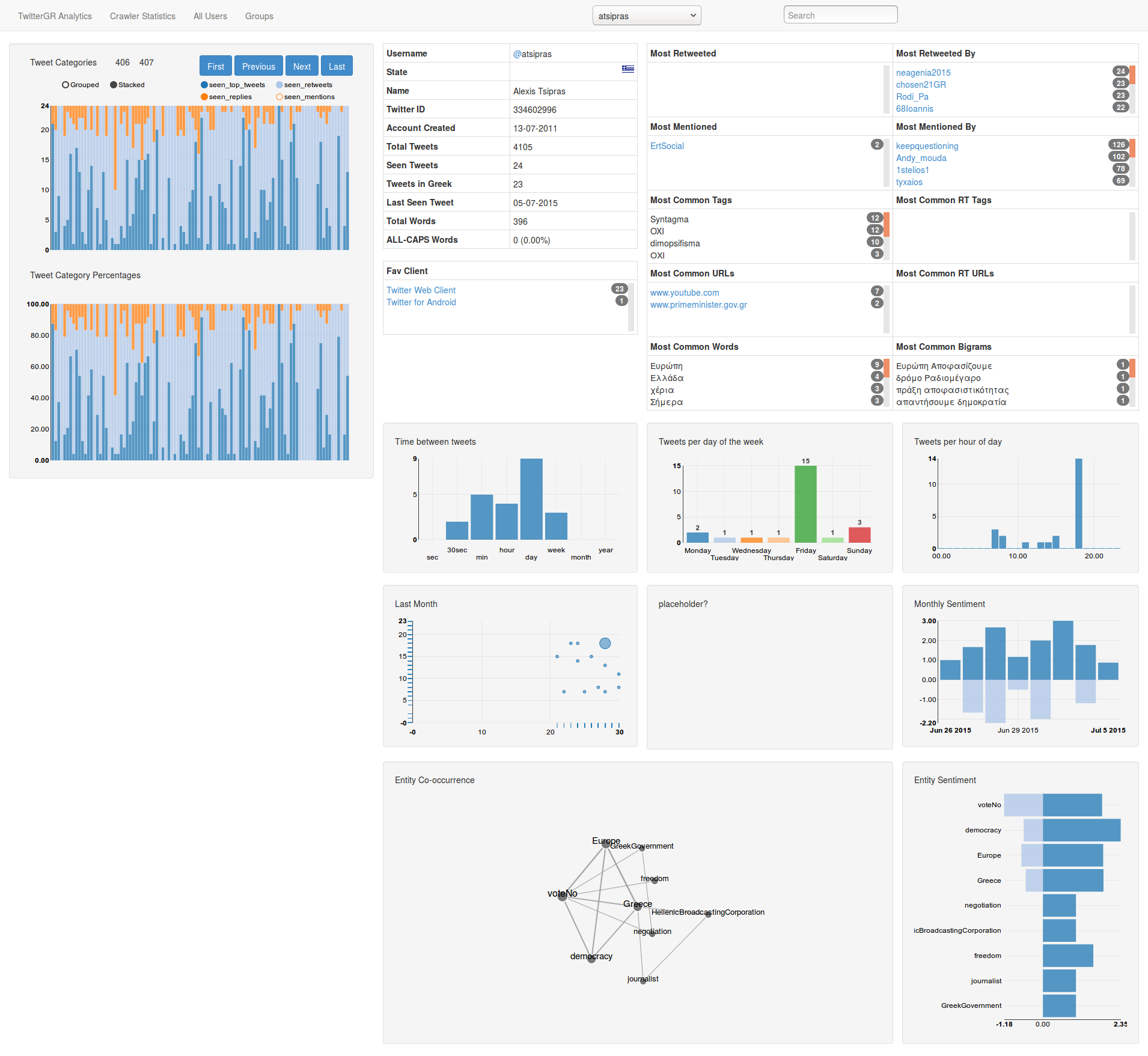}
\caption{Dashboard application for user vector analysis}
\label{fig:dashboard}
\end{figure}

\begin{figure*}
\noindent
\resizebox{0.49\textwidth}{!}{
\begin{tabular}{|p{0.21\textwidth}|p{0.40\textwidth}|}
\hline
{\feat id} & The twitter User ID. \\
{\feat screen\_name} & The user's screen name. If the user has more than one screen name, we use the last seen screen name. \\
{\feat screen\_name\_len} & Length of the user's screen name~\cite{pennacchiotti2011machine}. \\
{\feat screen\_name\_upper} & Number of uppercase letters in the user's screen name~\cite{pennacchiotti2011machine}. \\
{\feat screen\_name\_lower} & Number of lowercase letters in the user's screen name~\cite{pennacchiotti2011machine}. \\
{\feat screen\_name\_digit} & Number of digits in the users' screen name~\cite{pennacchiotti2011machine}. \\
{\feat screen\_name\_alpha} & Number of letters in the users' screen name~\cite{pennacchiotti2011machine}. \\
{\feat name} & The name of the user. \\
{\feat name\_len} & Length of the user's name. If the user has used many names over time, this is the last seen name used by the user~\cite{pennacchiotti2011machine}. \\
{\feat name\_upper} & Number of uppercase letters in the user's name~\cite{pennacchiotti2011machine}. \\
{\feat name\_lower} & Number of lowercase letters in the user's name~\cite{pennacchiotti2011machine}. \\
{\feat name\_digit} & Number of digits in the user's name~\cite{pennacchiotti2011machine}. \\
{\feat name\_alpha} & Number of letters in the user's name~\cite{pennacchiotti2011machine}. \\
{\feat name\_greek} & Number of greek letters in the user's name. \\
{\feat created\_at} & The date the user first joined twitter~\cite{pennacchiotti2011machine}. \\
{\feat tweet\_count} & Total tweet count, as reported by twitter~\cite{pennacchiotti2011machine}. \\
{\feat favourites\_count} & Number of tweets this user has favorited, as reported by twitter~\cite{uddin2014understanding}. \\
{\feat followers\_count} & Number of users that follow this user, as reported by twitter~\cite{pennacchiotti2011machine}. \\
{\feat friends\_count} & Number of users this user follows, as reported by twitter~\cite{pennacchiotti2011machine}. \\
{\feat fr\_fo\_ratio} & Ratio of friends to followers~\cite{pennacchiotti2011machine}. \\
{\feat location} & User's location as reported by the user~\cite{pennacchiotti2011machine}. \\
{\feat has\_location} & Presence of the location field~\cite{pennacchiotti2011machine}. \\
{\feat time\_zone} & User's time zone as reported by the user. \\
{\feat lang} & The language this user selected for the twitter UI. \\
{\feat protected} & Was the user protected the last crawl time. \\
{\feat verified} & True if this is a verified user~\cite{uddin2014understanding}. \\
{\feat dead} & True if this account was seen to be deleted. \\
{\feat suspended} & True if this account was suspended the last time this user was crawled. \\
{\feat user\_url} & URL field of the user's profile, if any. \\
{\feat bio\_words} & Number of words in the user's profile. \\
{\feat bio\_upper\_words} & Number of all-uppercase words in the user's bio. \\
{\feat bio\_lower\_words} & Number of all-lowercase words in the user's bio. \\
{\feat bio\_punctuation\_chars} & Number of punctuation characters in the user's bio. \\
{\feat bio\_digit\_chars} & Number of digits in the user's profile. \\
{\feat bio\_alpha\_chars} & Number of letters in the user's profile. \\
{\feat bio\_upper\_chars} & Number of uppercase letters in the user's bio. \\
{\feat bio\_lower\_chars} & Number of lowercase letters in the user's bio. \\
{\feat bio\_greek\_chars} & Number of greek letters in the user's bio. \\
{\feat bio\_total\_chars} & Length of the user's profile. \\
{\feat seen\_total} & Total number of this user's tweets used in computing this vector. \\
{\feat total\_inferred} & Total number of this user's tweets including encountered tweet IDs that could not be crawled and are probably deleted. \\
{\feat seen\_greek\_total} & Total number of this user's tweets marked as being in Greek by the twitter API. \\
{\feat all\_intervals} & A histogram of time-between-tweets for all tweets seen. \\
{\feat seen\_top\_tweets} & Number of seen tweets that were not retweets or replies. \\
{\feat top\_tweets\_pcnt} & Percentage of seen tweets that were not retweets or replies. \\
{\feat top\_intervals} & A histogram of time-between-tweets for all seen tweets that were not retweets or replies. \\

{\feat mention\_indegree} & Number of users that mention this user. \\
{\feat mention\_outdegree} & Number of users mentioned by this user. \\
{\feat mention\_inweight} & Number of tweets that mention this user. \\
{\feat mention\_outweight} & Number of mentions by this user~\cite{uddin2014understanding}. \\
{\feat mention\_avg\_inweight} & Average mentions of this user per mentioner. \\
{\feat mention\_avg\_outweight} & Average mentions per mentioned user. \\
{\feat mention\_out\_in\_ratio} & Out-degree/in-degree ratio (mention reciprocation). \\
{\feat mention\_pcnt} & Percentage of seen tweets that are mentions. \\
{\feat most\_mentioned\_users} & A list of the users most mentioned by this user, and the mention counts. \\
{\feat most\_mentioned\_by} & Users that are seen to most mention this user, and the mention counts. \\
{\feat retweet\_indegree} & Number of users that have retweeted this user. \\
{\feat retweet\_outdegree} & Number of users this user retweeted. \\
{\feat retweet\_inweight} & Number of retweets of this user's tweets. \\
\hline
\end{tabular}
} \hfil
\resizebox{0.49\textwidth}{!}{
\begin{tabular}{|p{0.21\textwidth}|p{0.40\textwidth}|}
\hline
{\feat retweet\_outweight} & Number of retweets by this user~\cite{uddin2014understanding}. \\
{\feat retweet\_avg\_inweight} & Average retweets per retweeter. \\
{\feat retweet\_avg\_outweight} & Average retweets per retweeted user. \\
{\feat retweet\_out\_in\_ratio} & Out-degree/in-degree ratio (retweet reciprocation). \\
{\feat retweet\_pcnt} & Percentage of seen tweets by this user that were retweets. \\
{\feat most\_retweeted\_users} & A list of the users most retweeted by this user, and the retweet counts. \\
{\feat most\_retweeted\_by} & A list of the users that most retweeted this user, and the retweet counts. \\
{\feat rt\_intervals} & A histogram of time-between-retweets by this user. \\

{\feat reply\_indegree} & Number of users that have replied to this user at least once~\cite{cheng2011predicting}. \\
{\feat reply\_outdegree} & Number of users to which this user replied at least once~\cite{cheng2011predicting}. \\
{\feat reply\_inweight} & Number of replies this user received~\cite{cheng2011predicting}. \\
{\feat reply\_outweight} & Number of replies tweeted by this user~\cite{uddin2014understanding,cheng2011predicting}. \\
{\feat reply\_avg\_inweight} & In-Replies per in-degree~\cite{cheng2011predicting}. \\
{\feat reply\_avg\_outweight} & Out-Replies per out-degree~\cite{cheng2011predicting}. \\
{\feat reply\_out\_in\_ratio} & Out-degree/in-degree ratio (replies sent for each received)~\cite{cheng2011predicting}. \\
{\feat replies\_pcnt} & Percentage of seen tweets by this user that were replies. \\
{\feat most\_replied\_to} & Users to whom this user most replied, and tweet counts. \\
{\feat most\_replied\_by} & Users that most replied to this user, and tweet counts. \\
{\feat reply\_intervals} & A histogram of time-between-replies by this user. \\
{\feat seen\_replied\_to} & Number of tweets that received replies from other users~\cite{uddin2014understanding}. \\
{\feat most\_engaging\_tweet} & The tweet by this user that gathered most replies by other users. \\

{\feat plain\_tweets} & Number of tweets without hashtags, mentions, or URLs~\cite{uddin2014understanding}. \\
{\feat most\_used\_sources} & List of twitter application clients used by this user and tweet counts per client, as reported by twitter. \\
{\feat time\_between\_any} & Minimum, maximum, average, median, and standard deviation of times between any seen tweets~\cite{pennacchiotti2011machine}. \\
{\feat time\_between\_top} & Minimum, maximum, average, median, and standard deviation of times between seen tweets that are not retweets or mentions. \\
{\feat time\_between\_rt} & Minimum, maximum, average, median, and standard deviation of times between seen tweets that are retweets. \\
{\feat time\_between\_replies} & A histogram of time-between-replies by this user. \\
{\feat max\_daily\_interval} &  Minimum, maximum, average, median, and standard deviation of the largest interval between actions in any given day~\cite{ferraz2015rsc}. \\
{\feat last\_tweeted\_at} & Time and date of the last seen tweet by this user. \\
{\feat life\_time} & Time from account creation to last seen tweet~\cite{uddin2014understanding}. \\
{\feat tweets\_per\_hour\_of\_day} & Histogram of total seen tweets by this user for each hour of day~\cite{pennacchiotti2011machine}. \\
{\feat tweets\_per\_weekday} & Histogram of seen tweets by this user for each day of the week. \\
{\feat tweets\_per\_active\_day} & Minimum, maximum, average, median, and standard deviation of tweets per active day, and days of minimum and maximum. \\
{\feat tweets\_per\_day} & Minimum, maximum, average, median, and standard deviation of tweets for every day from creation to now, and days of minimum and maximum. \\
{\feat last\_month} & A histogram of tweet counts seen during the last month of this user's seen activity, per hour of day. \\

{\feat fr\_scanned\_at} & Last time this user's list of friends was crawled. \\
{\feat seen\_fr} & Total number of twitter users seen to be followed by this user at least once. \\
{\feat gr\_fr} & Number of friends that are marked as Greek-speaking. \\
{\feat gr\_fr\_pcnt} & Number of friends that are marked as Greek-speaking. \\
{\feat tr\_fr} & Number of friends tracked. \\
{\feat tr\_fr\_pcnt} & Percentage of friends tracked. \\
{\feat fo\_scanned\_at} & Last time this user's list of followers was crawled. \\
{\feat seen\_fo} & Total number of twitter users seen to follow this user at least once. \\
{\feat gr\_fo} & Number of followers marked as Greek-speaking. \\
{\feat gr\_fo\_pcnt} & Number of followers marked as Greek-speaking. \\
{\feat tr\_fo} & Number of followers currently tracked. \\
\hline
\end{tabular}
}
\caption{List of features per user (1/2)}
\label{tab:features1}
\end{figure*}

\begin{figure*}
\noindent
\resizebox{0.49\textwidth}{!}{
\begin{tabular}{|p{0.21\textwidth}|p{0.40\textwidth}|}
\hline
{\feat tr\_fo\_pcnt} & Percentage of followers currently tracked. \\
{\feat fr\_fo\_jaccard} & Jaccard similarity between the friend and follower sets. \\
{\feat fr\_and\_fo} & Size of the intersection of the friend and follower sets, \emph{i.e.}, all reciprocal follow edges. \\
{\feat fr\_or\_fo} & Size of the union of the friend and follower sets, \emph{i.e.}, all neighbors in the follow graph. \\
{\feat gr\_fr\_fo} & Number of all friends and followers marked as Greek-speaking. \\
{\feat gr\_fr\_fo\_pcnt} & Percent of Greek-speaking neighbors (friends and followers). \\
{\feat greek} & True if this user was inferred to be Greek or Greek-speaking. \\
{\feat total\_words} & Total number of words in all seen tweets. \\
{\feat min\_wptw} & Minimum number of words per seen tweet. \\
{\feat avg\_wptw} & Average number of words per seen tweet. \\
{\feat med\_wptw} & Median number of words per seen tweet. \\
{\feat std\_wptw} & Standard deviation of the number of words per tweet. \\
{\feat unique\_words} & Number of unique words that this user has used in seen tweets. \\
{\feat lex\_freq} & Lexical frequency (the ratio of unique words over total words) over all seen tweets. \\
{\feat total\_bigrams} & Number of bigrams constructed from seen tweets. \\
{\feat unique\_bigrams} & Number of unique bigrams in seen tweets. \\
{\feat bigram\_lex\_freq} & Lexical frequency (the ratio of unique bigrams over total bigrams) over all seen tweets. \\
{\feat articles} & Number of times this user was seen using an article (article list mined from Greek Wiktionary) \\
{\feat pronouns} & Number of times this user was seen using a pronoun (pronoun list mined from Greek Wiktionary) \\
{\feat expletives} & Number of times this user was seen using an expletive (expletive list mined from Greek Wiktionary) \\
{\feat locations} & Number of times this user was seen using the name of a place (location list mined from Greek Wiktionary) \\
{\feat emoticons} & Number of times this user was seen using an emoticon. \\
{\feat emoji} & Number of times this user was seen using an emoji. \\
{\feat alltokens} & All tokens of text seen in this user's tweets (words, hashtags, mentions, emoticons, etc.) \\
{\feat all\_caps\_words} & Number of words seen to be in all-capital letters, excluding words of length 1. \\
{\feat all\_caps\_words\_pcnt} & Percentage of words seen to be in all-capital letters excluding words of length 1. \\
{\feat all\_caps\_tweets} & Number of tweets seen to be in all-capital letters. \\
{\feat all\_caps\_tweets\_pcnt} & Percentage of seen tweets to be in all-capital letters. \\
{\feat all\_nocaps\_words} & Number of words seen to have no capital letters. \\
{\feat all\_nocaps\_words\_pcnt} & Percentage of words seen to have no capital letters. \\
{\feat punctuation\_chars} & Number of characters that are punctuation in this user's seen tweets. \\
{\feat punctuation\_pcnt} & Percent of characters that are punctuation in this user's seen tweets. \\
{\feat total\_chars} & Total number of characters in seen tweets. \\
{\feat digit\_chars} & Total number of digits in seen tweets. \\
{\feat digit\_pcnt} & Percentage of characters in seen tweets that were digits. \\
{\feat alpha\_chars} & Total number of letters in this user's seen tweets. \\
{\feat alpha\_pcnt} & Percentage of letters in seen tweets. \\
{\feat upper\_chars} & Total number of uppercase letters in seen tweets. \\
\hline
\end{tabular}
} \hfil
\resizebox{0.49\textwidth}{!}{
\begin{tabular}{|p{0.21\textwidth}|p{0.40\textwidth}|}
\hline
{\feat upper\_pcnt} & Percentage of uppercase letters in seen tweets. \\
{\feat lower\_chars} & Total number of lowercase letters in this user's seen tweets. \\
{\feat lower\_pcnt} & Percentage of characters in seen tweets that were lowercase letters. \\
{\feat greek\_chars} & Total number of greek letters in this user's seen tweets. \\
{\feat greek\_pcnt} & Percentage of characters in seen tweets that were greek letters. \\
{\feat total\_hashtags} & Total number of hashtags this user's tweets~\cite{uddin2014understanding}. \\
{\feat hashtags\_per\_tw} & Minimum, maximum, average, median, and standard deviation of hashtags per tweet~\cite{pennacchiotti2011machine,uddin2014understanding}. \\
{\feat uniq\_hashtags} & Number of unique hashtags seen in tweets written by this user. \\
{\feat total\_rt\_hashtags} & Total number of hashtags in tweets retweeted by this user. \\
{\feat uniq\_rt\_hashtags} & Number of unique hashtags seen in tweets retweeted by this user. \\
{\feat most\_common\_words} & List of most common words (excluding stop-words) used by this user, and their counts. \\
{\feat most\_common\_bigrams} & List of most common bigrams (excluding stop-words) used by this user, and their counts. \\
{\feat most\_common\_hashtags} & List of most common hashtags used by this user, and their counts. \\
{\feat most\_common\_\newline \hbox{\hspace{1.9cm} rt\_hashtags}} & List of most common hashtags in tweets retweeted by this user, and their counts. \\
{\feat most\_common\_urls} & List of most common domain names in URLs found in this user's tweets, and their counts. \\
{\feat most\_common\_rt\_urls} & List of most common domain names in URLs found tweets retweeted by this user, and their counts. \\
{\feat seen\_urls} & Total number of URLs in all tweets~\cite{uddin2014understanding}. \\
{\feat urls\_per\_tw} & Minimum, maximum, average, median, and standard deviation of URLs per tweet~\cite{pennacchiotti2011machine,uddin2014understanding}. \\
{\feat avg\_edit\_distance} & Average edit distance between every posted URL and profile name~\cite{uddin2014understanding}. \\
{\feat daily\_sentiment} & Average positive and negative sentiment per tweet mentioning each sentiment, per day, as timeseries~\cite{antonakaki:greferendum}. \\
{\feat entity\_overlap} & A graph of entities. Node weights are counts of tweets mentioning each entity, edge weights are counts of tweets mentioning both entities. \\
{\feat senti\_entities} & List of average positive and negative sentiment scores for all seen tweets mentioning an entity, per entity. \\
{\feat favoriters} & Number of users seen to have liked a tweet by this user.\\
{\feat favorited} & Number of users whose any tweet this user has liked.\\
{\feat most\_favoriters} & A list of users that have liked this user's tweets the most, and the number of likes per user.\\
{\feat most\_favorited} & A list of users whose tweets this user has liked the most, and the number of likes per user.\\
{\feat lexical\_gender} & A struct of two numbers: Percentages of self references that are done using the male and female gender of the word. \\
{\feat number\_of\_languages} & Number of languages used by this user~\cite{subrahmanian2016darpa}. \\
{\feat tweets\_per\_language} & Number of tweets for the five languages most used. \\
{\feat vector\_timestamp} & The UTC timestamp of this vector (this is used by the crawler engine for caching user vectors). \\
\hline
\end{tabular}
}
\caption{List of features per user (2/2)}
\label{tab:features2}
\end{figure*}

\begin{figure*}
\centering
\begin{tabular}{|l|c|c|c|c|c|}
\hline
Graph    &      $|V|$ &         $|E|$ & Size & Directed & Weighted \\
\hline
Follow   & 26,339,971 &   204,969,957 & 4,2G &      Yes &  No \\
Retweet  &  3,851,055 &    46,084,224 & 1,0G &      Yes & Yes \\
Mention  &  2,226,118 &     2,781,915 &  65M &      Yes & Yes \\
Reply    &  4,552,175 &    24,364,103 & 552M &      Yes & Yes \\
Quote    &  1,279,360 &     4,794,911 & 112M &      Yes & Yes \\
List Sim & 54,309,000 & 1,993,937,542 &  44G &      No  & Yes \\
Favorite &  2,778,775 &    38,881,162 & 893M &      Yes & Yes \\
\hline
\end{tabular}
\caption{Graphs mined from data}
\label{fig:graphs}
\end{figure*}

\begin{figure*}
\centering
\begin{minipage}{0.33\textwidth}
\includegraphics[width=\textwidth]{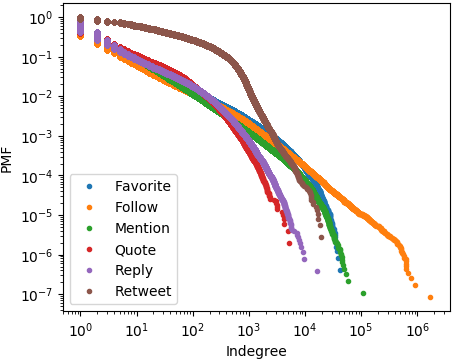}
\caption{Distribution of in-degree}
\label{fig:indegree}
\end{minipage}%\hfill
%\end{figure}
%
%\begin{figure}
\begin{minipage}{0.33\textwidth}
\includegraphics[width=\textwidth]{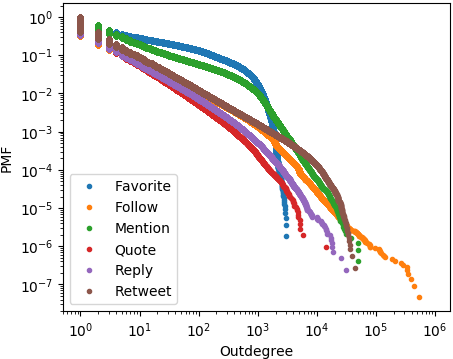}
\caption{Distribution of out-degree}
\label{fig:outdegree}
\end{minipage}%\hfill
%\end{figure}
%
%\begin{figure}
\begin{minipage}{0.33\textwidth}
\includegraphics[width=\textwidth]{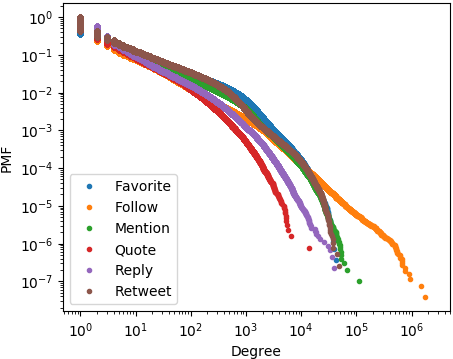}
\caption{Distribution of degree}
\label{fig:degree}
\end{minipage}
\end{figure*}

\begin{figure*}
\centering
\includegraphics[width=0.33\textwidth]{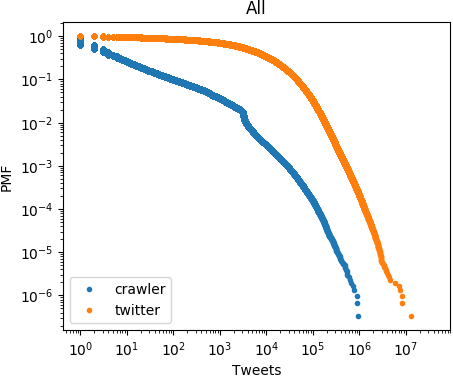}
%\hfill
\includegraphics[width=0.33\textwidth]{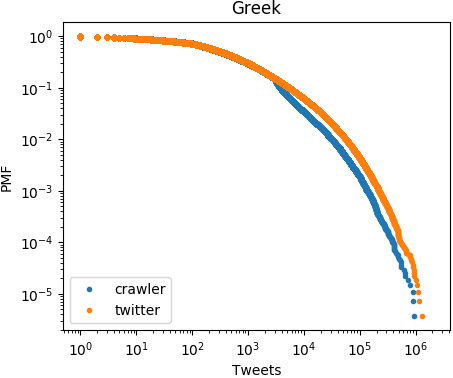}
\caption{Distribution of tweets per user}
\label{fig:twperuser}
\end{figure*}

In addition to user discovery and tracking, \tool is able to mine the
crawled data and produce a large set of features for every twitter
user, to facilitate subsequent analysis.  Tables~\ref{tab:features1}
and~\ref{tab:features2} shows a list of user features produced.  Many
of these features have been previously documented and used in related
work on various classification or generalization use cases.  In
implementing these analyses out-of-the-box, \tool provides a framework
for rapid prototyping of new research and rapid replication of
existing work.  The features include all information provided per-user
by the twitter API, as well as features extracted from the tweets of
each user, from the metadata of the tweets, and from a user's
relations to other users.  For use cases where it is important to
follow the evolution of the data, \tool includes tools to compute
time-interval versions of all features for specific users or groups of
users.
\tool includes a web-based dashboard, depicted in
Figure~\ref{fig:dashboard}, to assist with analysis of crawled tweets.
The dashboard depicts a selection of the mined features, including
time series of a user's sentiment per day, sentiment per entity, as
well as neighbor nodes for multiple relations.

\section{Data Processing and Graph Extraction}

To assist with social network analytics applications, \tool performs
additional post-processing of the harvested data to generate a set of
graphs.  In addition to the Follow Graph, stored as a dynamic graph of
timestamped edges, \tool also scans tweets to generate a set of graphs
among users, namely the Retweet Graph, Mention Graph, Reply Graph,
Quote Graph; all are directed and weighted, where each edge's weight
is the number of observed interactions of the corresponding type, from
the source user to the target user.  Moreover, \tool mines the list
and list membership data crawled to generate the List Similarity
Graph; it is an undirected graph where an edge between two users
amounts to membership of the same list, and an edge's weight is the
number of lists that include both users.  Finally, using the crawled
favorites per user \tool extracts the user-to-user Favorite Graph.

Figure~\ref{fig:graphs} shows the size of the mined graphs in vertices
and edges, their size on disk, as well as whether they are directed
or weighted.
Figures~\ref{fig:indegree}, \ref{fig:outdegree} and \ref{fig:degree} presents the in-degree, out-degree, and
undirected degree for the interaction graphs.  The list similarity
graph is not included, as it could not be easily analyzed by \tool
using a single machine.  Note that the directionality of the graphs
follows action, which in the case of retweets is not the direction
information flows. That is, an edge in the Follow Graph points from
the follower to the followee, an edge in the Retweet Graph points from
the retweeter to the tweeter, an edge in the Mention Graph points from
the mentioner to the mentioned, an edge in the Reply Graph points from
the replier to the replied-to, an edge in the Quote Graph points from
the quoter to the quoted, and an edge in the Favorite Graph points
from the favoriter to the favorited.

%\section{Coverage Evaluation}

To evaluate the coverage \tool achieves for the crawled users,
Figure~\ref{fig:twperuser} compares the distribution of tweets per user
as reported by twitter in the user information returned for each
user, with the total count of tweets crawled and saved by \tool.
Even though \tool worked for a brief duration compared to the active
period of most users, it was able to discover and crawl a very large
percentage of the tweets of even the most prolific users.  Part of
this is possible for active users because \tool follows retweets,
replies, and quotes to the past and discovers very old tweets that are
not otherwise reachable using the standard API.

\begin{figure}
\centering
\includegraphics[width=0.42\textwidth]{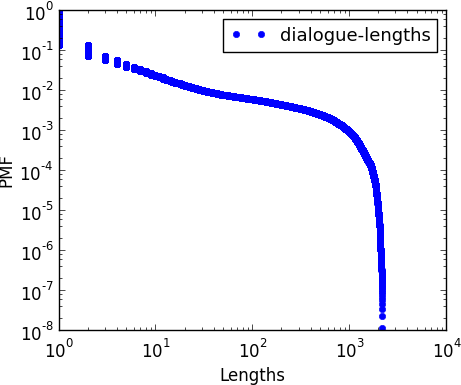}
\caption{Distribution of thread lengths}
\label{fig:dialogues}
\end{figure}

As a simple use case for \tool's usability, Figure~\ref{fig:dialogues}
shows a simple computation of the maximum lengths for all threads of
replies found starting with tweets by users marked as Greek-speaking.
The distribution is extremely skewed, with 86\% consisting of a tweet
and a single reply and 6.3\% consisting of two replies, while the
single longest thread has length 2185.

\section{Related Work}

%\subsection{Crawling}
There is a lot of related research focusing on the twitter social
network; this section presents representative samples of such work and
discusses how \tool compares or advances the state of the art.

TwitterEcho~\cite{twitterecho} is a Twitter crawler focused on
discovering and crawling small communities.  The authors design a
targeted crawler and use it to recognize and track Portuguese
accounts.  TwitterEcho also prioritizes account crawling by ordering
users according to their activity patterns, opting to crawl active
users more frequently.  TwitterEcho is a distributed system requiring
multiple clients to crawl and aggregating the results into a single
server. \tool  can also be used in this fashion for large
communities, but we found that for communities having on the order of
100,000 to 200,000 active users like the Greek-speaking twitter users,
one computer suffices.  TwitterEcho collected more than 14 million
tweets from 100,000 users within 11 months.  \tool manages to
use many more available endpoints from the twitter REST API, crawling,
in addition to tweets and user information, the follow graph, the
favorites of all tracked users, as well as list ownership,
subscription, and membership information.  \tool uses the fact that
the Greek alphabet suffices to recognize the language, while
TwitterEcho resorts to a more complex language classifier to separate
Portuguese from Brazilian.

%\subsection{Graph Computations}
%
%Backstrom \emph{et al.}~\cite{backstrom2006group} study the evolution
%an growth of communities in two online social networks, DBLP and
%LiveJournal.  They use a set of per-community features and relations
%between communities and individual nodes, and find that community
%growth depends not only on proximity but also on community density,
%and the density of each node's neighborhood.
%
%Cattuto \emph{et al.}~\cite{cattuto2013time} use the Neo4J graph
%database to analyze a time-dependent social graph generated using
%wearable sensors that captured in-person meetings of all the
%participants.  They were able to execute several complex queries with
%a temporal component over the graph and report scalability issues as
%the representation of the dynamic graph (as a static graph) tends to
%become very dense over time.
%
%Cheng \emph{et al.}~\cite{cheng2011predicting} study the twitter
%mention graph and predict reciprocity of communication, using a
%combined approach of decision trees and regression models, on a wide
%set of graph features extracted for the neighborhood of each vertex.
%They find social status difference to be a very important predictor of
%reciprocity.

%\subsection{Classification}

There is a very large body of literature on feature extraction from
twitter content, most of which uses the features to perform
classification.  \tool is currently able to efficiently extract a very
large subset of all the features mentioned in the following papers.
Sakaki \emph{et al.}~\cite{sakaki2014twitter} augment gender classification
with image processing, classifying images as one of cartoon/illustration,
famous person, food, goods, memo/leaflet, outdoor/nature, person, pet,
screenshot/capture, and other.
They use a "bot detection" criterion of whether a user has less than 150 tweets
made using a mobile or web client.
Liu and Ruths~\cite{liu2013s} use first name, as reported by the user, for
gender inference. We expect that this feature is highly language-dependent, and
will perform even better in the Greek language, where names, nouns, etc, are
gendered. 
Zhang \emph{et al.}~\cite{zhang2016your} use content and interaction
features to construct a model for classifying twitter users into age groups.
Uddin \emph{et al.}~\cite{uddin2014understanding} use a wide set of
features extracted from user information to classify users into six
categories: personal users, professional users, business users, spam
bots, news feed bots, and viral marketing bots.
Hu \emph{et al.}~\cite{hu2017language} correlate twitter and linkedin
data to classify users into the categories: marketing, administrator,
start-up, editor, software engineer, public relations, office clerk,
or designer.  The authors perform sentiment analysis and present
sentiment results on the Pearson scale for each class.
Pennacchiotti and Popescu present two versions of the same
work~\cite{pennacchiotti2011machine,pennacchiotti2011democrats} in
which they use features in four classes to classify users, namely user
profile, user tweeting behavior, linguistic content of user messages,
and user social network features; and employ a set of hand crafted
regular expressions to mine ethinicty and gender.%
%\footnote{The two papers are very similar, share more than a few paragraphs of
%text, do not cite each other, and must have been under review concurrently for
%KDD 2011 (submission on February 18) and ICWSM 2011 (submission on February 7),
%which is quite impressive, to say the least.}
%They infer ethnicity and gender using a set of regular expressions that match
%frequent relevant phrases.  They construct a trained probabilistic model using
%a few \emph{seed users} per class to find \emph{proto words} and associate
%words or n-grams with user classes with probabilities (similar to
%LDA~\cite{blei2003latent}).  Then they select the k-most important proto-words
%per class and use them to generate the linguistic features.  \tool
%does not currently use lexical features like this as it would require
%knowing the classes in advance and manual training with selected
%``ground truth'' users.
%They do similar analyis for selecting important hashtags per class and using
%them to construct features.  They also use LDA over all users (on 20000 words
%per user) to generate a set of 100 most important ``generic'' topics. Then they
%generate a feature value per selected topic.  This is repeated for
%domain-specific topics, relevant to only the training set of users, per class.
%
DARPA has organized a competition on bot
detection~\cite{subrahmanian2016darpa}.
%The challenge specifically targets
%influence bots above other bots, \emph{i.e.}, bots that aim to diffuse a
%specific sentiment on a specific topic.
All teams used an array of features, where the winning team managed to
create visualizations that assisted in rapidly recognizing bots.  The
competition report describes multiple machine-learning techniques used
to cluster users and find bot outliers, detect bot-to-bot
similarities, etc.

%Ferrara \emph{et al.}~\cite{ferrara2016predicting} use a set of user,
%time and network features to predict interaction with extremist
%content promoters.
%
%Volkova and Bell~\cite{volkova2017identifying} use a set of content, network,
%sentiment and image features to predict accounts that are to be deleted or
%suspended.
%
%
%Bakshy \emph{et al.}~\cite{bakshy2011everyone} measure twitter user influence
%and diffusion patterns of URLs.

The study of user activity over time has been studied in previous
work, showing that extraction of timeseries data from content may
offer valuable insights into user behavior.
\tool mines several timeseries from user activity, including
daily sentiment per entity or in total, idle time intervals, etc.,
facilitating further experimentation in that direction.
Paraskevopoulos \emph{et al.}~\cite{paraskevopoulos2016tweet} use the
time series of twitter activity to correlate users without location
information, with already geotagged users.
Ferraz \emph{et al.}~\cite{ferraz2015rsc} study inter-action time
intervals in twitter activities and create a theoretical model that
closely explains and reproduces observations.  They use their model to
discover outliers and detect bots.
Bild \emph{et al.}~\cite{bild2015aggregate} focus on the Retweet Graph
and reason about the effects of sampling on a set of metrics such as
the distributions of tweets per user, tweet rates, and inter-event
time intervals. They find that the Retweet Graph is small-world and
scale-free similarly to the follow graph, but with stronger
clustering.

%Geolocating users could also be done by correlating with localized events, such
%as concerts, earthquakes, disasters, accidents, etc.
%TODO: FIND RELATED WORK, OR DO IT.

Twitter lists have been used as crowdsourced similarity metrics in the
past, interpreting the fact that twitter users independently
classify other users into their own lists.  \tool extracts and
processes list membership information and enables multiple similar use
cases to be explored.
Kim \emph{et al.}~\cite{kim2010analysis} use twitter lists to
recognize representative words for all of the list traffic and
associate these words with list members.  They also mine keywords from
the list names as representative for the list members, even if these
words are not used by the members.
Wu \emph{et al.}~\cite{wu2011says} use Twitter lists to recognize
elite from ordinary users and study the usage patterns and content
posted in each class.  They use twitter lists to recognize elite users
via their list-similarity with exemplar accounts.
Culotta and Cutler~\cite{culotta2016mining} present a method for
computing brand perception as a set of metrics of similarity to
Entities.  They mine entity definitions using twitter lists to
identify characteristic accounts, and use Jaccard similarity on
follower sets to compute a distance metric from chosen entities.

%
%\subsection{Political}
%
%O'Connor \emph{et al.}~\cite{oconnor2010tweets} correlate aggregate tweet
%sentiment (computed with english sentiment lexicon) with poll results.
%They state that temporal smoothing is critically important for noise
%reduction in the sentiment time series. They find up to 0.8 correlation of
%twitter sentiment with poll results.
%
%Charalampakis \emph{et al.}~\cite{charalampakis2016comparison} use a set of
%lexical features to predict irony in political tweets and show it correlates to
%election results.
%

Often it is useful to analyze content in a non-user-centric way, to
observe propagation patterns or orchestrated behavior, as is the case
in bot detection work.  Although \tool performs user-centric crawling
and aggregation of data, it includes tools to extract subsets of the
reply, quote, mention, and retweet graphs conditional on time
intervals or specific content, that allow detailed monitoring of
information propagation as used in related work.
Ratkiewicz \emph{et al.}~\cite{ratkiewicz2011detecting} search for
astroturfing, or orchestrated campaigns appearing to be grass-roots movements
in order to influence opinion. They use hashtags, mentions, URLs and the entire
text of every tweet as ``memes'' and look into propagation patterns in
diffusion networks.  They use a set of features per meme to classify memes,
%including the
%number of nodes involved,
%number of propagation edges involved,
%mean node degree,
%mean node strength,
%mean edge weight in the largest connected component,
%maximum in, out degree,
%user with maximum in, out strength,
%maximum in, out strength,
%standard deviation of in, out degree,
%standard deviation of in, out strength,
%skew of in, out degree distributions,
%skew of in, out strength distributions,
%mean size of connected components,
%size of largest connected component,
%number of unique injections of the meme,
%number of times the `truthy' button was clicked for this meme by human users surveyed,
and assign six GPOMS sentiment dimensions~\cite{bollen2011twitter}: calm, alert,
sure, vital, kind, happy.
Tsur and Rappoport~\cite{tsur2015don} analyze use of hashtags in
tweets and create a model that predicts the popularity of hashtags
based on features like length, capitalization, abbreviations, and
number of keystrokes required to type.
Anderson \emph{et al.}~\cite{anderson2012effects} study user
similarity metrics to predict evaluations and election results. They
capture both content similarity and similarity of user interactions,
and find that relative social status affects how similarity influences
user opinions.

%Jungherr~\cite{jungherr2016twitter} presents a systematic literature review of
%papers studying twitter influence in elections.

Previous work has focused on greek-speaking twitter users, although at
a much smaller scale, and focusing on tweets related to specific
events, containing specific keywords or hashtags, etc. In comparison,
\tool targets users instead of tweets, allowing researchers to study
specific events in the proper context of existing user relations and
interactions.
Antonakaki \emph{et al.}~\cite{antonakaki:greferendum} present an analysis of
the Greek 2015 referendum and parliamentary elections.  They use a stemmer for
word matching, and a lexicon-based sentiment analysis to assign sentiment to
LDA topics.  They produced an accurate prediction of the ballot results by
counting number of tweets and tweet sentiment per topic, and assigning topics
to outcomes.  \tool uses the same sentiment dictionary, kindly
provided by the authors, to extract sentiment features per user and
per entity.
%
%Bond \emph{et al.}~\cite{bond201261} describe a large experiment that
%took place on Facebook during the 2010 elections in the USA, where
%they measure the network effect of pro-voting advertising.  They find
%very large effect reaching to friends-of-friends in the social graph,
%very much more prominent for ``strong'' edges (edges with lots of user
%interactions) relative to ``weak'' edges (edges without interactions).
%
%Morales \emph{et al.}~\cite{morales2015measuring} use a physical model
%of bipole charge to model political polarization, and demonstrate is
%use on a dataset from Venezuelan twitter traffic. They analyze daily
%traffic and generate a time-evolving model of polarization, finding
%that a small minority of elite users influence most of the observed
%polarization.
%
Theocharis \emph{et al.}~\cite{theocharis2015using} analyze twitter
content related to social movements and classify tweets into
categories according to their intent. They find a small part of tweets
are related to actions and organization, with most being reports about
the events and conversations.

\section{Conclusions}

\tool is a lightweight, user-centric crawler that targets
twitter communities based on language.  It satisfies twitter crawler
constraints and does not require using multiple crawling accounts.
Furthermore, it includes post-processing primitives that facilitate
data analysis especially with respect to mining relationship graphs,
as well as extraction of data in a quasi-anonymized (user id-only)
form that can be shared or published without violating twitter's terms
of use.  \tool can assist SNA researchers in gathering data from
twitter without violation of its terms.  It also facilitates the quick
testing of hypotheses or quick replication of previous work by
including a wide set of primitives and common computations,
out-of-the-box.  Although \tool aims to be lightweight and can be
easily deployed on a single machine, further analysis of the data may,
however, require more resources.

\bibliographystyle{IEEEtran}
\bibliography{polyvios}

\begin{thebibliography}{10}
\providecommand{\url}[1]{#1}
\csname url@samestyle\endcsname
\providecommand{\newblock}{\relax}
\providecommand{\bibinfo}[2]{#2}
\providecommand{\BIBentrySTDinterwordspacing}{\spaceskip=0pt\relax}
\providecommand{\BIBentryALTinterwordstretchfactor}{4}
\providecommand{\BIBentryALTinterwordspacing}{\spaceskip=\fontdimen2\font plus
\BIBentryALTinterwordstretchfactor\fontdimen3\font minus
  \fontdimen4\font\relax}
\providecommand{\BIBforeignlanguage}[2]{{%
\expandafter\ifx\csname l@#1\endcsname\relax
\typeout{** WARNING: IEEEtran.bst: No hyphenation pattern has been}%
\typeout{** loaded for the language `#1'. Using the pattern for}%
\typeout{** the default language instead.}%
\else
\language=\csname l@#1\endcsname
\fi
#2}}
\providecommand{\BIBdecl}{\relax}
\BIBdecl

\bibitem{pennacchiotti2011machine}
M.~Pennacchiotti and A.-M. Popescu, ``A machine learning approach to twitter
  user classification.'' \emph{Icwsm}, vol.~11, no.~1, pp. 281--288, 2011.

\bibitem{uddin2014understanding}
M.~M. Uddin, M.~Imran, and H.~Sajjad, ``Understanding types of users on
  twitter,'' \emph{arXiv preprint arXiv:1406.1335}, 2014.

\bibitem{cheng2011predicting}
J.~Cheng, D.~M. Romero, B.~Meeder, and J.~Kleinberg, ``Predicting reciprocity
  in social networks,'' in \emph{Privacy, Security, Risk and Trust (PASSAT) and
  2011 IEEE Third Inernational Conference on Social Computing (SocialCom), 2011
  IEEE Third International Conference on}.\hskip 1em plus 0.5em minus
  0.4em\relax IEEE, 2011, pp. 49--56.

\bibitem{ferraz2015rsc}
A.~Ferraz~Costa, Y.~Yamaguchi, A.~Juci Machado~Traina, C.~Traina~Jr, and
  C.~Faloutsos, ``{RSC}: Mining and modeling temporal activity in social
  media,'' in \emph{Proceedings of the 21th ACM SIGKDD International Conference
  on Knowledge Discovery and Data Mining}.\hskip 1em plus 0.5em minus
  0.4em\relax ACM, 2015, pp. 269--278.

\bibitem{antonakaki:greferendum}
D.~Antonakaki, D.~Spiliotopoulos, C.~V. Samaras, S.~Ioannidis, and
  P.~Fragopoulou, ``Investigating the complete corpus of referendum and
  elections tweets,'' in \emph{2016 IEEE/ACM International Conference on
  Advances in Social Networks Analysis and Mining (ASONAM)}, Aug 2016, pp.
  100--105.

\bibitem{subrahmanian2016darpa}
V.~Subrahmanian, A.~Azaria, S.~Durst, V.~Kagan, A.~Galstyan, K.~Lerman, L.~Zhu,
  E.~Ferrara, A.~Flammini, and F.~Menczer, ``The darpa twitter bot challenge,''
  \emph{Computer}, vol.~49, no.~6, pp. 38--46, 2016.

\bibitem{twitterecho}
M.~Bo{\v{s}}njak, E.~Oliveira, J.~Martins, E.~Mendes~Rodrigues, and
  L.~Sarmento, ``{TwitterEcho}: a distributed focused crawler to support open
  research with twitter data,'' in \emph{Proceedings of the 21st International
  Conference on World Wide Web}.\hskip 1em plus 0.5em minus 0.4em\relax ACM,
  2012, pp. 1233--1240.

\bibitem{sakaki2014twitter}
S.~Sakaki, Y.~Miura, X.~Ma, K.~Hattori, and T.~Ohkuma, ``Twitter user gender
  inference using combined analysis of text and image processing,'' \emph{V\&L
  Net}, vol. 2014, p.~54, 2014.

\bibitem{liu2013s}
W.~Liu and D.~Ruths, ``What's in a name? using first names as features for
  gender inference in twitter.'' in \emph{AAAI spring symposium: Analyzing
  microtext}, vol.~13, 2013, p.~01.

\bibitem{zhang2016your}
J.~Zhang, X.~Hu, Y.~Zhang, and H.~Liu, ``Your age is no secret: Inferring
  microbloggers' ages via content and interaction analysis.'' in \emph{ICWSM},
  2016, pp. 476--485.

\bibitem{hu2017language}
T.~Hu, H.~Xiao, T.-v.~T. Nguyen, and J.~Luo, ``What the language you tweet says
  about your occupation,'' \emph{ICWSM}, 2016.

\bibitem{pennacchiotti2011democrats}
M.~Pennacchiotti and A.-M. Popescu, ``Democrats, republicans and starbucks
  afficionados: user classification in twitter,'' in \emph{Proceedings of the
  17th ACM SIGKDD international conference on Knowledge discovery and data
  mining}.\hskip 1em plus 0.5em minus 0.4em\relax ACM, 2011, pp. 430--438.

\bibitem{paraskevopoulos2016tweet}
P.~Paraskevopoulos, G.~Pellegrini, and T.~Palpanas, ``When a tweet finds its
  place: fine-grained tweet geolocalisation,'' in \emph{International workshop
  on data science for social good (SoGood), in conjunction with the European
  conference on machine learning and principles and practice of knowledge
  discovery (ECML PKDD)}, 2016.

\bibitem{bild2015aggregate}
D.~R. Bild, Y.~Liu, R.~P. Dick, Z.~M. Mao, and D.~S. Wallach, ``Aggregate
  characterization of user behavior in twitter and analysis of the retweet
  graph,'' \emph{ACM Transactions on Internet Technology (TOIT)}, vol.~15,
  no.~1, p.~4, 2015.

\bibitem{kim2010analysis}
D.~Kim, Y.~Jo, I.-C. Moon, and A.~Oh, ``Analysis of twitter lists as a
  potential source for discovering latent characteristics of users,'' in
  \emph{ACM CHI workshop on microblogging}, 2010, p.~4.

\bibitem{wu2011says}
S.~Wu, J.~M. Hofman, W.~A. Mason, and D.~J. Watts, ``Who says what to whom on
  twitter,'' in \emph{Proceedings of the 20th international conference on World
  wide web}.\hskip 1em plus 0.5em minus 0.4em\relax ACM, 2011, pp. 705--714.

\bibitem{culotta2016mining}
A.~Culotta and J.~Cutler, ``Mining brand perceptions from twitter social
  networks,'' \emph{Marketing Science}, vol.~35, no.~3, pp. 343--362, 2016.

\bibitem{ratkiewicz2011detecting}
J.~Ratkiewicz, M.~Conover, M.~R. Meiss, B.~Gon{\c{c}}alves, A.~Flammini, and
  F.~Menczer, ``Detecting and tracking political abuse in social media.''
  \emph{ICWSM}, vol.~11, pp. 297--304, 2011.

\bibitem{bollen2011twitter}
J.~Bollen, H.~Mao, and X.~Zeng, ``Twitter mood predicts the stock market,''
  \emph{Journal of computational science}, vol.~2, no.~1, pp. 1--8, 2011.

\bibitem{tsur2015don}
O.~Tsur and A.~Rappoport, ``Don't let me be\# misunderstood: Linguistically
  motivated algorithm for predicting the popularity of textual memes.'' in
  \emph{ICWSM}, 2015, p. 426.

\bibitem{anderson2012effects}
A.~Anderson, D.~Huttenlocher, J.~Kleinberg, and J.~Leskovec, ``Effects of user
  similarity in social media,'' in \emph{Proceedings of the fifth ACM
  international conference on Web search and data mining}.\hskip 1em plus 0.5em
  minus 0.4em\relax ACM, 2012, pp. 703--712.

\bibitem{theocharis2015using}
Y.~Theocharis, W.~Lowe, J.~W. Van~Deth, and G.~Garc{\'\i}a-Albacete, ``Using
  twitter to mobilize protest action: online mobilization patterns and action
  repertoires in the occupy wall street, indignados, and aganaktismenoi
  movements,'' \emph{Information, Communication \& Society}, vol.~18, no.~2,
  pp. 202--220, 2015.

\end{thebibliography}

\end{document}